\documentclass[11pt]{article}
\usepackage{geometry}                
\usepackage{amsmath}
\usepackage{amscd}
\usepackage{times}
\usepackage{epsfig}
\usepackage{amsfonts,amsthm,eucal,amssymb,amscd}
\usepackage{color}
\usepackage{graphicx}
\usepackage{epstopdf}
\usepackage{authblk}
\title{Polynomial symmetries of spherical Lissajous systems}

\author[1]{J.A. Calzada}
\author[2]{\c{S}. Kuru}
\author[3]{J. Negro}
\affil[1]{Departamento Matem\'atica Aplicada, Universidad de
Valladolid, 47011 Valladolid, Spain} \affil[2]{Department of
Physics, Faculty of Sciences, Ankara University, 06100 Ankara,
Turkey} \affil[3]{Departamento de F\'{\i}sica Te\'orica, At\'omica y
\'Optica, Universidad de Valladolid\\47071 Valladolid, Spain}
\begin{document}
\maketitle

\begin{abstract}
In a previous work, both the constants of motion of a classical system
and the symmetries of the corresponding quantum version have been computed
with the help of factorizations.
As their expressions were not polynomial,
in this paper the question of finding  an equivalent set of polynomial
constants of motion and symmetries is addressed. The general algebraic
relations as well as the appropriate Hermitian relations will also be found.
\end{abstract}




\section{Introduction}

In a former paper \cite{calzada14}, henceforth referred as
Lissajous--1, we have characterized a type of symmetries for a
particular classical system that we called `Lissajous system on the
sphere'. The name `Lissajous' for this system (and many others that
share this feature) comes from the form of its bounded trajectories:
they are closed and similar to the usual Lissajous curves, where the
motion in the variables have a rational frequency. We have found
what we will call a set of `fundamental constants of motion'
(sometimes they will be referred as classical symmetries) given by
four functions $\langle H,H_\phi,X^\pm\rangle $, where $H$ and
$H_\phi$ are trivial symmetries due the separation of the two
variables, while $X^\pm$ is a pair of complex nontrivial symmetries.
In total there are three independent symmetries, but they are not
polynomial since $X^\pm$ include square roots of some functions.
Such roots are well defined for any physical motion of the system.

In the same paper, we have considered the corresponding quantum
Lissajous system on the sphere. We have computed another set of
`fundamental symmetries' that consist of four operators $\langle
\hat H, \hat H_\phi, \hat X^\pm \rangle$, which are in
correspondence with the above classical symmetries. Again, some
square roots of operators are present, but such expressions are well
defined in the physical space generated by eigenfunctions.

We have applied a general unifying method that was introduced in
Ref.\ \cite{celeghini13} to compute the fundamental symmetries of
classical and quantum systems. This method is based on the well
known factorizations of quantum mechanics
\cite{schrodinger40,infeldhull51} and a classical version called
`classical factorizations'  \cite{sengul08}.

Now, in this paper we will show that the above mentioned fundamental
symmetries determine in a straightforward way the polynomial
symmetries of both classical and quantum systems. Therefore, in this
way we can eliminate the somewhat `problematic' square roots and
state our results in the frame of polynomial superintegrable
systems. Besides, we will easily get the algebraic structure that
such polynomial symmetries will close, as well as their Hermitian
properties.

Section 2 will be devoted to polynomial symmetries of the classical
system, while the next one will implement the same approach to get
the polynomial symmetries of the quantum system. We will end with
some conclusions, in particular, we will also comment our results in
the light of other previous contributions in the literature on the
symmetries of this type of systems. We have included all the basic
results of Lissajous--1, so that the present paper is
self--contained and can be read independently.
%

\section{The classical system}

We will consider a classical system on the sphere (of unit radius) whose
Hamiltonian is
\begin{equation}\label{ch2a}
H = p_{\theta}^2 +\frac{1}{\sin^2 {\theta}}\,
p^2_{\varphi}+\frac{k^2 \alpha^2}{\sin^2 {\theta}\cos^2{k \varphi}}\, ,
\end{equation}
where $0<\theta<\pi$ and $-\pi/(2k)<\varphi<\pi/(2k)$.
In spherical coordinates,
the first two terms correspond to the kinetic part of the system,
while the last one is for the potential.
We will assume that
$k\geq 1/2$, in order that this system be well defined on the sphere.
After a change of canonical variables $\phi = k \varphi$, $p_\phi = p_\varphi/k$
we get an equivalent system given by the Hamiltonian
\begin{equation}\label{ch2b}
H = p_{\theta}^2 +\frac{k^2}{\sin^2 {\theta}}\left(
p^2_{\phi}+\frac{\alpha^2}{\cos^2{\phi}}\right)\, ,
\end{equation}
where now the range of the variable $\phi$ is $-\pi/2<\phi<\pi/2$.
\subsection{Ladder and shift functions}
This Hamiltonian is explicitly separated in the variables $\theta$
and $\phi$, so that two constants of motion will be identified with
two one--dimensional Hamiltonians,
\begin{equation}\label{cpth}
H _{\phi}=p^2_{\phi}+\frac{\alpha^2}{\cos^2{\phi}}\,
\end{equation}
and
\begin{equation}\label{cht1}
H_{\theta}^M=p_{\theta}^2+\frac{k^2\,H_\phi}{\sin^2
{\theta}}=p_{\theta}^2+\frac{M^2}{\sin^2 {\theta}}\,,\qquad M= k^2 H_\phi\, .
\end{equation}
In order to build  nontrivial symmetries we need two types of functions related
to the above two Hamiltonians.

The first pair of functions $B^\pm(\phi,p_\phi)$ are called ladder functions
for $H_\phi$.
They are given by \cite{sengul08}
\begin{equation}\label{cbpm}
B^{\pm}=\mp i \cos{\phi}\,p _{\phi}+\sqrt{H_{\phi}}\,\sin{\phi} \,.
\end{equation}
For a real motion where $H_\phi> 0$, they are complex conjugate functions.
These ladder functions factorize
the Hamiltonian in the form
\begin{equation}\label{cpth2}
H _{\phi}=B^+\,B^-+\alpha^2\,.
\end{equation}
The set  $\langle H _{\phi},B^\pm \rangle$ satisfy the
following PBs,
\begin{equation}\label{commpt}
\{H _{\phi},B^{\pm}\}=\mp\,2\, i\,\sqrt{H _{\phi}}\,B^{\pm},
\qquad \{B^-,B^+\} = -2i\sqrt{H_\phi} \,.
\end{equation}

The second set of functions $A^\pm(\theta,p_\theta,M)$ are called shift functions of $H_{\theta}^M$.
They are also formed by two complex conjugate functions
(for a real motion) given by
\begin{equation}\label{capm}
A^{\pm}=\mp i\,p_{\theta}+M \cot{\theta} \,.
\end{equation}
In this case, they factorize the Hamiltonian $H_{\theta}^M$ in the  form
\begin{equation}\label{cht2}
H_{\theta}^M=
A^{+}\,A^{-}+M^2 \, .
\end{equation}
The functions $A^{\pm}$ and $H_{\theta}^M$ obey the following PBs
\begin{equation}\label{commpt3}
\{H _{\theta}^M,A^{\pm}\}=\pm 2i\,\frac{M}{\sin^2{\theta}}\,A^{\pm}\,,\qquad
\{A^-,A^+\} = 2i\frac{M}{\sin^2\theta} 
\, .
\end{equation}
\subsection{The algebra of fundamental and polynomial symmetries}

If the parameter $k$ that appear in the initial Hamiltonian (\ref{ch2b}) is rational,
$k = m/n$, where $m,n\in \mathbb N$,
then a pair of symmetries $X^\pm$, such that $\{H,X^\pm\}=0$, is built in terms of
the above sets of ladder and shift functions \cite{calzada14},
\begin{equation}\label{csymmet1}
X^{\pm}=(B^{\pm})^n(A^{\pm})^m\,.
\end{equation}
Therefore, at this stage we have  four `fundamental symmetries'
$\langle H, H_\phi, X^\pm \rangle$ whose constant values will be
denoted by $E,E_\phi,Q^\pm$, respectively. In fact, as $X^\pm$ are
complex conjugate functions,  we can write $Q^\pm = q\, e^{\pm i\,
\phi_0}$, where $q$ is for the modulus and $\phi_0$ is for the
phase. The last two constants of motion are not independent since,
due to the factorizations (\ref{cpth2}) and (\ref{cht2}), their
product is
\begin{equation}\label{prod}
X^+X^- = (E_\phi-\alpha^2)^n(E-k^2E_\phi)^m\,, \qquad E_\phi\geq \alpha^2\,,
E\geq k^2E_\phi \, .
\end{equation}
The algebraic structure of the fundamental symmetry set
$\langle H, H_\phi, X^\pm \rangle$ can be obtained by computing its PBs in a direct
way from (\ref{commpt}) and (\ref{commpt3}):
\begin{equation}\label{pbfs}
\left\{\begin{array}{l}
\{H_\phi, X^\pm\} = \mp 2i\, n \sqrt{H_\phi} \, X^\pm,
\qquad \{H, H_\phi\} = \{H,X^\pm\} = 0
\\[2.ex]
\{X^+,X^-\} = 2i\sqrt{H_\phi}
\left[-km^2(H_\phi{-}\alpha^2) {+}n^2(H{-}k^2H_\phi)\right](H_\phi{-}\alpha^2)^{n{-}1}
(H {-}k^2H_\phi)^{m{-}1} \, .
\end{array}\right.
\end{equation}

\noindent As we know from Lissajous--1, the symmetries $X^\pm$ are
quite appropriate to describe the trajectories of the motion for
this system. In particular, we have shown that they are a kind of
Lissajous curves on the sphere. However, this class of symmetries
are not polynomial, due to the square root that takes part in the
functions $A^\pm$ and $B^\pm$, so we will show below how they can
supply us the relevant polynomial symmetries.

Let us compute explicitly the expression of $X^\pm$ by expanding the powers in (\ref{csymmet1}),
\begin{equation}\label{exp}
X^{\pm}=(\mp i \cos{\phi}\,p _{\phi}+\sqrt{H_{\phi}}\,\sin{\phi} )^n
(\mp i\,p_{\theta}+ k \sqrt{H_{\phi}} \cot{\theta})^m\,.
\end{equation}
Depending on the parity type of $m+n$ we will write the result according
to the following convention.

\begin{itemize}
\item[(a)]
For $m+n$ even we use the notation,
\begin{equation} \label{even}
X^{\pm} = \pm i \,{\cal O} \sqrt{H_\phi} + {\cal E}
\end{equation}

\item[(b)]
For $m+n$ odd we write the result in the form,
\begin{equation} \label{odd}
X^{\pm} =  {\cal O} \sqrt{H_\phi} \mp i\, {\cal E} \, .
\end{equation}
\end{itemize}
The functions ${\cal O}(\phi,\theta,p_\phi,p_\theta)$ and ${\cal E}(\phi,\theta,p_\phi,p_\theta)$
are real polynomial functions of $p_\phi,p_\theta$; the first
one comes from odd  and the second from the even degrees of $\sqrt{H_\phi}$ in the
expansion (\ref{exp}). In both cases the polynomials
(in $p_\phi,p_\theta$), ${\cal O}$ and ${\cal E}$ are of degree $n+m-1$ and $n+m$, respectively.
These two polynomial functions are symmetries of the initial Hamiltonian,
\begin{equation}
\{H,{\cal O}\} = \{H,{\cal E}\} = 0\, .
\end{equation}
In conclusion, we have obtained an equivalent set of symmetries that
will be called `polynomial symmetries',
\begin{equation}\label{ps}
\langle H,H_\phi, {\cal O}, {\cal E} \rangle \, .
\end{equation}
Its algebraic structure can be computed by inserting the expressions (\ref{even})
or (\ref{odd}) in (\ref{pbfs}).
Finally, we arrive at the following non--vanishing PBs
\begin{equation}\label{pbps}
\left\{
\begin{array}{l}
\{H_\phi, {\cal O}\} = - 2n\,{\cal E},\quad \{H_\phi, {\cal E}\} =  2n\, H_\phi {\cal O}
\\[2.ex]
\{{\cal O},{\cal E}\} = -n\, {\cal O}^2 +
\left[ -km^2(H_\phi-\alpha^2) +n^2(H -k^2\, H_\phi)\right](H_\phi-\alpha^2)^{n-1}(H -k^2\, H_\phi)^{m-1} \, .
\end{array}\right.
\end{equation}
The dependence relation of the four polynomial generators is obtained from (\ref{prod}),
\begin{equation}\label{prod2}
{\cal O}^2 H_\phi + {\cal E}^2= (H_\phi-\alpha^2)^n(H-k^2H_\phi)^m\, .
\end{equation}
In other words, the polynomial symmetries close a polynomial algebra
(with respect to the generators $H,H_\phi,{\cal O}, {\cal E}$),
where all the square roots  have been eliminated in the final
expressions. For the case $k=1$, this polynomial algebra will be
quadratic, for other cases, the degree will be $m+n-1$.

\subsection{Examples of classical polynomial symmetry algebras}
We will give some examples to illustrate the general results of the previous subsections.
In this way we can see how are the concrete expressions of polynomial algebras for
some simple cases. We have checked the following formulas with the help of
{\sl Mathematica}.

\begin{itemize}
\item
Case $m=1$, $n=1$, $k=1$
\begin{equation}\label{o11}
{\cal O} = -p_\phi \cot \theta \cos \phi - \sin \phi \,p_\theta
\end{equation}
\begin{equation}\label{e11}
{\cal E} = -\cos\phi\,p_\theta\,p_\phi +
 \cot\theta (p_\phi^2 \sin\phi + \alpha^2 \sec\phi \tan\
\phi)
\end{equation}
\begin{equation}\label{pbps11}
\begin{array}{l}
\{H_\phi, {\cal O}\} = - 2\,{\cal E},\qquad \{H_\phi, {\cal E}\} =
2\, H_\phi {\cal O}
\\[2.ex]
\{{\cal O},{\cal E}\} = -\, {\cal O}^2
 - (H_\phi-\alpha^2) +(H -H_\phi)
\end{array}
\end{equation}

\item
Case $m=1$, $n=2$, $k=1/2$
\begin{equation}\label{o12}
{\cal O} = \frac{1}{2} \left[-p_\phi^2 \cos^2\phi \cot\theta -
   4\,p_\theta\, p_\phi \,\cos\phi \sin\phi+
   \cot\theta \left(p_\phi^2 \sin^2\phi + \alpha^2 \tan^2\phi\right)\right]
\end{equation}
\begin{equation}\label{e12}
{\cal E} = -p_\theta\, p_\phi^2 \cos^2\phi +
 p_\phi^3 \cos\phi \cot\theta \sin\phi +
 p_\theta\, p_\phi^2 \sin^2\phi + \alpha^2 \tan\phi \left(p_\phi \cot\theta + p_\theta
 \tan\phi\right)\,
\end{equation}
\begin{equation}\label{pbps12}
\begin{array}{l}
\{H_\phi, {\cal O}\} = - 4\,{\cal E},\qquad \{H_\phi, {\cal E}\} =
4\, H_\phi {\cal O}
\\[2.ex]
\{{\cal O},{\cal E}\} = -2\,{\cal O}^2+ \left[-\frac{1}{2} (H_\phi -
\alpha^2) + 4 (H - \frac{1}{4} H_\phi)\right] (H_\phi - \alpha^2)
\end{array}
\end{equation}

\item
Case $m=2$, $n=1$, $k=2$
\begin{equation}\label{o21}
{\cal O} = -4\,p_\theta\, p_\phi \cos\phi \cot\theta -
\left(p_\theta^2 -
    4\,p_\phi^2 \cot^2\theta\right) \sin\phi+
 4 \alpha^2 \cot^2\theta \sec\phi \tan\phi
\end{equation}
\begin{equation}\label{e21}
{\cal E} =-p_\phi \cos\phi \left(p_\theta^2 - 4\,p_\phi^2
\cot^2\theta\right) +
 4\,\cot\theta \left[p_\phi \,\alpha^2 \cot\theta\,\sec\phi +
    p_\theta\, p_\phi^2 \sin\phi +
    p_\theta\, \alpha^2 \sec\phi\,\tan\phi\right]\,
\end{equation}
\begin{equation}\label{pbps21}
\begin{array}{l}
\{H_\phi, {\cal O}\} = - 2\,{\cal E},\qquad \{H_\phi, {\cal E}\} =
2\, H_\phi {\cal O}
\\[2.ex]
\{{\cal O},{\cal E}\} = -{\cal O}^2 + \left[ -8(H_\phi-\alpha^2) +(H
-4\,H_\phi)\right](H -4\,H_\phi)
\end{array}
\end{equation}

\item
Case $m=3$, $n=1$, $k=3$
\begin{equation}\label{o31}
\begin{array}{l}
{\cal O}=9 p_\phi \cos\phi \cot\theta \left(p_\theta^2 -
    3 p_\phi^2 \cot^2\theta\right) -
 27 p_\phi \alpha^2 \cot^3 \theta\sec\phi +
 p_\theta^3 \sin\phi
\\[1.5ex]
-27 p_\theta\, \cot^2\theta \left(p_\phi^2 \sin\phi + \alpha^2
\sec\phi \tan\phi\right)
 \end{array}
\end{equation}
\begin{equation}\label{e31}
\begin{array}{l}
{\cal E} =p_\theta\, p_\phi \cos\phi \left(p_\theta^2 - 27\,
p_\phi^2\, \cot^2\theta\right) - 9 \cot\theta [3 \,p_\theta\,
p_\phi\,\alpha^2 \cot\theta\,\sec\phi
\\[1.5ex]
-\frac{3 }{4}\left(p_\phi^2 {+} 2\,\alpha^2 {+} p_\phi^2 \,\cos
{2\phi}\right)^2 \cot^2\theta \sec^3\phi \tan\phi {+} p_\theta^2
\left(p_\phi^2 \sin\phi {+} \alpha^2 \sec\phi \tan{\phi}\right)]
\end{array}
\end{equation}
\begin{equation}\label{pbps31}
\begin{array}{l}
\{H_\phi, {\cal O}\} = - {\cal E},\qquad \{H_\phi, {\cal E}\} =
H_\phi {\cal O}
\\[2.ex]
\{{\cal O},{\cal E}\} = - {\cal O}^2 + \left[ -27(H_\phi-\alpha^2)
+(H -9\,H_\phi)\right](H -9\,H_\phi)^2
\end{array}
\end{equation}

\end{itemize}

\section{The quantum system}
In quantum mechanics the corresponding system is described by the Hamiltonian operator
in spherical coordinates (the units $2m=1$ and $\hbar=1$ have been chosen):
\begin{equation}\label{qh1}
\hat{H} = - \partial^2_{\theta} - \cot{\theta}\partial_{\theta}
-\frac{1}{\sin^2 {\theta}} \partial^2_{\varphi}
+\frac{k^2 \alpha^2}{\sin^2 {\theta} \cos^2{k\,\varphi}}
\end{equation}
where $0<\theta<\pi$ and  $0<\varphi<\pi/2k$. The first three terms describe the
kinetic energy, they are obtained from the Laplacian in spherical coordinates; the
last term is for the potential. After the change of variable $\phi = k\varphi$,
we get a new equivalent Hamiltonian that we will use hereafter,
\begin{equation}\label{qh2}
\hat{H} = - \partial^2_{\theta} - \cot{\theta}\partial_{\theta}
+\frac{k^2}{\sin^2 {\theta}}\left(
-\partial^2_{\phi}+\frac{\alpha^2}{\cos^2{\phi}}\right)\,.
\end{equation}
The corresponding eigenvalue equation
\begin{equation}\label{eve1}
\hat{H}\,\Psi(\theta,\phi)=E\,\Psi(\theta,\phi)\,,
\end{equation}
is a partial differential equation separated in the variables $\theta$ and $\phi$.
Then, we will look for separable solutions,
$\Psi(\theta,\phi)=\Theta_E^M(\theta)\Phi_\epsilon(\phi)$  characterized by
\begin{equation}\label{eve2}
\hat{H}^M_{\theta}\,\Theta_E^M(\theta)=E_{\theta}\,\Theta_E^M(\theta),
\qquad\hat{H}_{\phi}\,\Phi_\epsilon(\phi)=E_{\phi}\,\Phi_\epsilon(\phi),\qquad
E_{\theta}:= E,\quad E_{\phi}:=\epsilon^2\,,
\end{equation}
where
\begin{equation}\label{qhp2}
\hat{H}_{\phi} = - \partial^2_{\phi} +\frac{\alpha^2}{\cos^2
{\phi}}
\end{equation}
\begin{equation}\label{qht1}
\hat{H}^M_{\theta} = - \partial^2_{\theta} -
\cot{\theta}\partial_{\theta} +\frac1{\sin^2
{\theta}}\,{M^2}\,, \qquad M^2 := k^2 \epsilon^2 \, .
\end{equation}
The operators $\hat{H}_{\phi}$ and $\hat{H}^M_{\theta}$ can be
considered as one--dimensional component Hamiltonians of the total
Hamiltonian (\ref{qh2}).

\subsection{Ladder and shift operators}
Since the Hamiltonian is separable, we already have a symmetry $\hat
H_\phi$ associated to this separation. As in the classical case, in
order to get a nontrivial symmetry we need the ladder and shift
operators of the one--dimensional Hamiltonians.

The ladder operators for $\hat H_\phi$ are given by \cite{sengul09}
\begin{equation}\label{bpm}
\hat{B}_{\epsilon}^+=-\cos{\phi}\,\partial_{\phi}+\epsilon\,\sin{\phi},
\qquad\hat{B}_{\epsilon}^-=\cos{\phi}\,\partial_{\phi}+(\epsilon+1)\,\sin{\phi} \,.
\end{equation}
They satisfy the following factorization relation
\begin{equation}\label{fact1}
\hat{B}_{\epsilon}^-\hat{B}_{\epsilon}^+ =
\hat{B}_{\epsilon+1}^+\hat{B}_{\epsilon+1}^-=\epsilon(\epsilon+1) - \alpha^2
\end{equation}
and their action on
eigenfuncions is
\begin{equation}\label{bpmpsi}
\hat{B}^+\,\Phi_{\epsilon}:=\hat{B}_{\epsilon}^+\,\Phi_{\epsilon}\propto\Phi_{{\epsilon}+1},
\qquad\hat{B}^-\,\Phi_{\epsilon+1}:=\hat{B}_{\epsilon}^-\,\Phi_{\epsilon+1}\propto\Phi_{\epsilon}\,.
\end{equation}
Therefore, as they change the energy eigenvalues of eigenfunctions
of the same Hamiltonian,  they are called pure--ladder operators.
It is convenient to use the free--index notation for operators, but
at the same time, one must be careful with the rules (\ref{bpmpsi})
to act on eigenfunctions.

The shift operators for the Hamiltonian $\hat H^M_\theta$ (\ref{qht1}) are obtained by
the standard factorization method \cite{sengul09}
\begin{equation}\label{facqht2}
\hat{H}_{\theta}^M=\hat{A}^+_M\hat{A}^-_M+\lambda_M=\hat{A}^-_{M+1}\hat{A}^+_{M+1}+\lambda_{M+1},
\qquad\lambda_M=M(M-1)
\end{equation}
where
\begin{equation}\label{apm}
\hat{A}^+_M=-\partial_{\theta}+(M-1)\cot{\theta},
\qquad\hat{A}^-_M=\partial_{\theta}+M \cot{\theta}\,.
\end{equation}
The  operators $\hat{A}^\pm_M$ implement
the following intertwining rules  in the hierarchy (\ref{facqht2}) of Hamiltonians
$\{\hat{H}_{\theta} ^M\}$,
\begin{equation}\label{aint}
\hat{A}^-_M\hat{H}_{\theta} ^M=\hat{H}_{\theta} ^{M-1}\hat{A}^-_M,
\qquad\hat{A}^+_M\hat{H}_{\theta} ^{M-1}=\hat{H}_{\theta} ^M\hat{A}^+_M \, .
\end{equation}
These rules can be written in a shorter notation by eliminating
the subindex in $\hat{A}^\pm$, but taking care of their action.
The  operators $\hat A^\pm$ keep the energy $E_\theta$, but change
the parameter $M$, so they are called pure--shift operators.
\subsection{Sets of fundamental and polynomial symmetries}

As in the classical case if $k=m/n$, $m,n\in \mathbb N$, then a pair of symmetry
operators $\hat{X}^\pm$ is given by
\begin{equation}\label{symmpm2}
\hat{X}^{\pm}=(\hat{A}^{\pm})^{m}\,(\hat{B}^{\pm})^n \, .
\end{equation}
The set of four symmetries $\langle \hat H, \hat H_\phi, \hat X ^{\pm}\rangle$ is
called fundamental set. They are not independent; from the factorization properties
(\ref{fact1}) and (\ref{facqht2})
it can be seen that the products $\hat X ^{+}\hat X ^{-}$ and $\hat X ^{-} \hat X ^{+}$
are functions of the diagonal operators $\hat H, \hat H_\phi$. We can explicitly compute them,
\begin{equation}\label{xpm1}
\begin{array}{l}
\displaystyle  \hat X ^{+} \hat X ^{-} =
\prod_{r=1}^n\left[(\sqrt{\hat H_\phi}{-}r) (\sqrt{\hat H_\phi}{-}r{+}1)
{-} \alpha^2\right]
\prod_{p=1}^m\left[\hat H {-} (k\sqrt{\hat H_\phi}{-}p) (k \sqrt{\hat H_\phi}
{-}p{+}1)\right]
\\[2.ex]
\hskip1.5cm = P_1(\hat H,\hat H_\phi) - P_2(\hat H,\hat H_\phi)\sqrt{\hat H_\phi}
\end{array}
\end{equation}
\begin{equation}\label{xpm2}
\begin{array}{l}
\displaystyle \hat X ^{-} \hat X ^{+} =
\prod_{r=1}^n\left[(\sqrt{\hat H_\phi}{+}r) (\sqrt{\hat H_\phi}{+}r{-}1)
{-} \alpha^2\right]
\prod_{p=1}^m\left[\hat H {-} (k \sqrt{\hat H_\phi}{+}p)
(k \sqrt{\hat H_\phi}{+}p{-}1)\right]
\\[2.ex]
\displaystyle \hskip1.5cm = P_1(\hat H,\hat H_\phi) + P_2(\hat H,\hat H_\phi)\sqrt{\hat H_\phi}\, ,
\end{array}
\end{equation}
where the rhs.\ expressions are assumed to act on an eigenfunction of $\hat H$ and
$\hat H_\phi$ characterized by the eigenvalues $E$ and  $\sqrt{\hat H_\phi}= \epsilon$.
The functions $P_1(\hat H,\hat H_\phi)$ and $P_2(\hat H,\hat H_\phi)$ are polynomial
in $\hat H$ and $\hat H_\phi$.
The quantum expressions (\ref{xpm1}) and (\ref{xpm2}) can be compared
to the corresponding classical expression (\ref{prod}).


The symmetry operators $\hat X ^{\pm}$  act on the
simultaneous eigenfunctions of $\hat H$ and $\hat H_\phi$; its action will give another
common eigenfunction of both $\hat H$ and $\hat H_\phi$. Therefore, they explain the degeneracy
of each energy level and hence, the degeneracy of the spectrum of $\hat H$. We can easily
compute the commutation relations of the fundamental symmetry set,
\begin{equation}\label{crx}
[\hat H_\phi,\hat X^\pm] = \hat X^\pm(\pm 2n \sqrt{\hat H_\phi} + n^2),\qquad
[\hat X^+, \hat X^-] = -2 P_2(\hat H,\hat H_\phi) \sqrt{\hat H_\phi} \, .
\end{equation}
The last commutator can be obtained simply by taking the difference of
(\ref{xpm1}) and (\ref{xpm2}). If we add these two expressions we will get
the constrain relation
\begin{equation}\label{constrain}
\hat X^+ \hat X^-+ \hat X^- \hat X^+ = 2 P_1(\hat H,\hat H_\phi)\,.
\end{equation}

One may be tempted to discard the operators $\hat X ^{\pm}$ as
`true' symmetries since they depend on a formal square root operator
$\sqrt{\hat H_\phi}$. However,
 $\hat X ^{\pm}$ are well defined operators in the space of eigenfunctions of
 $\hat H$ (in fact, the operators $\hat X ^{\pm}$ are defined even in the space of formal
 eigenfunctions, physical or not, of the differential operator (\ref{qh2})).
We can use them in order to
 get polynomial symmetries that exclude square roots. To show this, let us expand
 expression (\ref{symmpm2}); here we will distinguish two cases.

 \begin{itemize}
 \item[(a)]
 If $m+n$ is even, the result of the expansion will be denoted by
 \begin{equation}\label{xpmeven}
 \hat X ^{\pm} = \pm \hat{\cal O} \sqrt{\hat H_\phi} + \hat{\cal E}
 \end{equation}
 In this expression $\hat{\cal O}$ and $\hat{\cal E}$ are  polynomial differential
 operators in $\partial_\phi, \partial_\theta$. Their degrees with respect to these
 partial derivatives are
 $m+n-1$, and $m+n$, respectively.


 \item[(b)]
 If $m+n$ is odd, then the resulting expansion will take the form
 \begin{equation}\label{xpmodd}
 \hat X ^{\pm} =  \hat{\cal O} \sqrt{\hat H_\phi} \pm \hat{\cal E} \,.
 \end{equation}
 In this case the polynomial $\hat{\cal O}$ and $\hat{\cal E}$ operators
 have the same degrees $m+n-1$ and  $m+n$.
 \end{itemize}

The explicit expressions of the polynomial differential operators $\hat{\cal O}$
and $\hat{\cal E}$ can be obtained in closed form, but this will not be needed
in what follows.
These operators are symmetries of the initial Hamiltonian (\ref{qh2}),
\begin{equation}\label{finite}
[\hat H, \hat{\cal O}] = [\hat H, \hat{\cal E}] =0 \,.
\end{equation}
Hence, we have arrived at a set of polynomial symmetries
\begin{equation}
\langle \hat H, \hat H_\phi, \hat{\cal O}, \hat{\cal E} \rangle \, .
\end{equation}

One may object that, strictly speaking, the finite differential
operators $\hat{\cal O}$ and  $\hat{\cal E}$ satisfy the symmetry equation
(\ref{finite}) only when it is applied to formal eigenfuntions
(physical or not) of the form
$\Psi= \Theta_E^M(\theta)\Phi_\epsilon(\phi)$, $M=k\epsilon$, that is,
 the set of functions
where the operators $\hat X^\pm$ are well defined (see
Lissajous--1). However, simple arguments show that if a linear
finite order partial differential operator in $\theta$ and $\phi$
annihilates the infinite linear space of all the formal
eigenfunctions $\Psi= \Theta_E^M(\theta)\Phi_\epsilon(\phi)$ then,
this operator must identically be zero.

The algebraic structure of the polynomial symmetries can be obtained from that
of the fundamental symmetries by
plugging  (\ref{xpm1}) and (\ref{xpm2}) into (\ref{crx}). The result is
\begin{equation}\label{pbpsb}
\left\{\begin{array}{l}
[\hat H_\phi, \hat {\cal O}] =
2n\,\hat {\cal E} + n^2 \hat {\cal O},
\quad [\hat H_\phi, \hat {\cal E}] =
2n\, \hat{\cal O} \hat H_\phi + n^2 \hat {\cal E}
\\[2.ex]
[\hat{\cal O},\hat {\cal E}] = -n\, \hat {\cal O}^2 \mp P_2(\hat H,\hat H_\phi)
\end{array}\right.
\end{equation}
where the signs $-$ and $+$ correspond to cases (a) and (b), respectively.
As we know by construction, these symmetries are not independent. The constrain relation,
obtained from (\ref{constrain}), has the form
\begin{equation}\label{restr1}
-\hat {\cal O}^2 \hat H_\phi - n\hat {\cal O}\hat{\cal E} +\hat{\cal E}^2
= \pm P_1(\hat H, \hat H_\phi)\,.
\end{equation}
Finally, we have arrived at a polynomial algebra of polynomial symmetries;
the degree of this algebra is the same as in the classical case: for $k=1$
the algebra is quadratic, and for $k\neq1$ the degree is $m+n-1$.

It is quite instructive to compare the formulas of the quantum
symmetry algebras of this subsection with those of Subsection 2.2
corresponding to  the classical algebras.

\subsection{Hermitian properties of polynomial symmetries}

Now, we will address the question of the Hermitian properties of the polynomial
symmetries. As the Hermitian properties of $\hat X^\pm$ in principle are not known,
we can not make use of the definition given in
(\ref{xpmeven}) and (\ref{xpmodd}) of $\hat {\cal O}$ and $\hat {\cal E}$ to find the
Hermitian conjugate of the polynomial operators. Therefore, in order to find them
we have to work with the following conditions:
\begin{itemize}
\item
The Hamiltonian $\hat H$ given in (\ref{qh2}) is Hermitian with respect to
the inner product
\begin{equation}
\langle \Psi_1,\Psi_2\rangle
= \int_{-\pi/2}^{\pi/2}d\phi\int_0^\pi d\theta\,
\Psi_1(\phi,\theta)^* \Psi_2(\phi,\theta) \sin \theta \, .
\end{equation}
So, the formal Hermitian conjugate of the differential operators are
\begin{equation}
(\partial_\theta)^\dagger = -(\partial_\theta + \cot \theta),\quad
(\partial_\phi)^\dagger = -\partial_\phi \,.
\end{equation}

\item
The Hermitian conjugate of a polynomial symmetry must be another polynomial
symmetry. As $\hat {\cal O}$ is the minimum order nontrivial symmetry,
we must have the following kind of Hermitian transformations:
\begin{equation}
\hat{\cal O}^\dagger = (-1)^{m+n-1}\hat{\cal O}+ Q(\hat
{H},\hat{H_\phi}),\qquad \hat{\cal E}^\dagger = (-1)^{m+n}\hat{\cal
E}  + \alpha \hat{\cal O} + R(\hat {H},\hat{H_\phi})
\end{equation}
where $\alpha$ is a constant, and the polynomials $Q(\hat
{H},\hat{H_\phi})$ and $R(\hat {H},\hat{H_\phi})$ must contribute in
lower degrees less than precedent terms.

\item
The Hermitian properties of  $\hat{\cal O}$ and $\hat{\cal E}$ must
be consistent with the commutation rules (\ref{pbpsb}).
\end{itemize}
We have found the following Hermitian rules that satisfy all the above requirements,
\begin{equation}
\hat{\cal O}^\dagger = (-1)^{m+n+1}\hat{\cal O},\qquad\hat {\cal
E}^\dagger = (-1)^{m+n}(\hat{\cal E} + n\, \hat{\cal O})\,.
\end{equation}
In the following examples, we have checked, that indeed the polynomial symmetries fulfill
the above Hermitian properties. We could change to an Hermitian/anti-Hermitian
basis if we introduce the new polynomial symmetry
\begin{equation}
\hat{\cal E}' = \hat{\cal E}+\frac{n}2\, \hat {\cal O}\,.
\end{equation}
The commutation relation in the new basis are
\begin{equation}\label{pbpsc}
\left\{\begin{array}{l}
[\hat H_\phi, \hat {\cal O}] =
2n\,\hat {\cal E}' \,,
\quad [\hat H_\phi, \hat {\cal E}'] =
n(\hat{\cal O} \hat H_\phi+ \hat H_\phi \hat{\cal O}) -\frac12 n^3 \hat {\cal O}
\\[2.ex]
[\hat{\cal O},\hat {\cal E}'] = -n\, \hat {\cal O}^2 \mp P_2(\hat H,\hat H_\phi)
\end{array}\right.
\end{equation}
and the restriction relation is
\begin{equation}\label{restr2}
-\hat {\cal O} \hat H_\phi \hat {\cal O} + \hat{\cal E}'{}^2
- \frac{n^2}4\hat {\cal O}^2 = \pm (P_1(\hat H, \hat H_\phi)
+ \frac{n}2 P_2(\hat H, \hat H_\phi)) \,.
\end{equation}
These new  relations are explicitly consistent with the Hermitian rules
\begin{equation}
{\hat{\cal O}}^\dagger = (-1)^{m+n-1}\hat{\cal O}\,,\qquad
\hat{{\cal E}'}^\dagger = (-1)^{m+n} \hat{\cal E}' \, .
\end{equation}

\subsection{Examples of quantum polynomial symmetry algebras}

Next we will show some examples for some values of $k$, just to see explicitly
a few simple realizations of these algebras.
\begin{itemize}
\item
Case $m=1$, $n=1$, $k=1$
\begin{align*}
    \hat{\cal O} =& -\cot \theta\,\cos\phi\,\partial_\phi-\sin \phi\,\,\partial_\theta \\[2.ex]
    \hat{\cal E} =&\,\cot \theta\,\sin \phi\,\hat{H}_\phi+\cos \phi\,\partial^{2}_{\theta\phi}\\[2.ex]
    P_1=&\,-\alpha^2\,\hat{H}+(\alpha^2-1+\hat{H})\hat{H}_\phi-\hat{H}^{2}_{\phi}   \\[2.ex]
    P_2=&\,\alpha^2+\hat{H}-2\,\hat{H}_\phi
\end{align*}
\item
Case $m=1$, $n=2$, $k=1/2$
\begin{align*}
    \hat{\cal O} =&\,\frac{1}{2}\,\cot \theta\left[\sin^2 \phi\,\hat{H}_\phi- \sin(2\,\phi)\,\partial_\phi
    +\cos^2 \phi\,\partial^{2}_{\phi}\right]+ \cos(2\,\phi)\,\partial_\theta+ \sin(2\,\phi)\,\partial^{2}_{\theta \phi}\\[2ex]
   \hat{\cal E} =&\, -\frac{1}{2}\,\cot \theta \left[\,\cos(2\,\phi)\,\hat{H}_\phi +
   \sin(2\,\phi)\,\partial_\phi\hat{H}_\phi \right]-\sin^2 \phi\,\partial_\theta \hat{H}_\phi+
   \sin(2\,\phi)\,\partial^{2}_{\theta\phi}\\
    -&\cos^2\phi\,\partial^{3}_{\theta\phi\phi}\\[2ex]
    P_1=&\, \alpha^2(\alpha^2-2)\,\hat{H}+\frac{1}{4}\left[-4+10\,\alpha^2-\alpha^4+4\,(5-2\,\alpha^2)\,\hat{H}\right]\,\hat{H}_\phi \\
&\,+ \frac{1}{4}\,(-13+2\,\alpha^2+4\,\hat{H})\,\hat{H}^{2}_{\phi}-\frac{1}{4}\,\hat{H}^{3}_{\phi}\\[2.ex]
P_2=&\,
\alpha^2\,(1-\frac{\alpha^2}{2})+2\,(1-2\,\alpha^2)\,\hat{H}+(2\,\alpha^2-3)\,\hat{H}_\phi
+4\,\hat{H}\,\hat{H}_\phi-\frac{3}{2}\,\hat{H}^{2}_{\phi}
\end{align*}
\item
Case $m=2$, $n=1$, $k=2$
\begin{align*}
   \hat{\cal O} =&\, -[3+\cos(2\,\theta)]\,\csc^2 \theta\,\cos \phi\, \partial_\phi+ 4\,\cot^2 \theta\,\sin \phi\,\hat{H}_\phi - \cot \theta (\sin \phi\,\partial_\theta\\
     -&\, 4\,\cos\phi\,\partial^{2}_{\theta\phi})+\sin \phi\, \partial^{2}_{\theta} \\[2.ex]
   \hat{\cal E} =&\, [3+2\,\cos(2\,\theta)]\,\csc^2 \theta\,\sin \phi\,\hat{H}_\phi- 4\, \cot^2\theta\,\cos \phi\, \partial_\phi\, \hat{H}_\phi - \cot \theta(4\,\sin \phi\, \partial_\theta\,\hat{H}_\phi\\
    -&\,\cos \phi\, \partial^{2}_{\theta\phi})- \cos \phi\,\partial^{3}_{\theta\theta\phi}\\[2.ex]
    P_1=&\, \alpha^2(2-\hat{H})\,\hat{H}+\left[4-20\,\alpha^2+(8\,\alpha^2-10)\,\hat{H}+\hat{H}^2\right]\,\hat{H}_\phi\\
    &\,+(52-16\,\alpha^2-8\,\hat{H})\,\hat{H}^{2}_{\phi}+16\,\hat{H}^{3}_{\phi}\\[2.ex]
    P_2=&-4\,\alpha^2+2\,(4\,\alpha^2-1)\,\hat{H}+\hat{H}^2+8\,(3-4\,\alpha^2\,-2\,\hat{H})\,\hat{H}_\phi+48\,\hat{H}^{2}_{\phi}
\end{align*}
\item
Case $m=3$, $n=1$, $k=3$
\begin{align*}
    \hat{\cal O} =&\, \frac{27}{2}\cot\theta \left[(3+\cos(2\,\theta))\csc^2 \theta\,\sin \phi\, \hat{H}_\phi -2\, \cos \phi\,\cot^2 \theta\,\partial_\phi\,\hat{H}_\phi-2\,\cot \theta\,\sin \phi\, \partial_\theta \,\hat{H}_\phi \right]\\
    -&\,\frac{3}{2}\left[15\,\cos \theta+ \cos(3\,\theta)\right]\,\csc^3 \theta\,\cos \phi\, \partial_\phi - [2+\cos(2\,\theta)]\csc^2 \theta\, \sin \phi\, \partial_\theta \\
    +&\,\cos \phi\,\left[18\,\cot^2 \theta+9\,\csc^2 \theta\right]\,\partial^{2}_{\theta\phi}+3\, \cot  \theta \sin \phi\,\partial^{2}_{\theta}- 9\,\cos\phi \cot \theta\,\partial^{3}_{\theta\theta\phi} - \sin \phi\, \partial^{3}_{\theta}    \\[2.ex]
   \hat{\cal E} =&\,27\,\cot^3 \theta\, \sin\phi\, \hat{H}^{2}_{\phi}+\frac{3}{2} \csc^2\theta \left[ (15\cos \theta + \cos(3\,\theta))\csc\theta\,\sin\phi\, \hat{H}_\phi \right.\\
    -&\, 3\left\{3\,(3+\cos(2\theta)) \cot\theta\,\cos\phi\, \partial_\phi\,\hat{H}_\phi- 6\, \cos^2\theta\, \cos\phi\, \partial_{\theta\phi}\,\hat{H}_\phi\right.\\
    +& \left. \left.\left.\sin\phi\left(2\,(2+\cos(2\,\theta))\,\partial_\theta\,\hat{H}_\phi -\sin(2\,\theta)\, \partial^{2}_{\theta}\,\hat{H}_\phi \right)\,\right\}\right]\right.\\
    +&\,\cos\phi\left[(2+\cos(2\,\theta))\csc^2\theta\,\partial^{2}_{\theta\phi} - 3\cot\theta \,\partial^{3}_{\theta\theta\phi}+\partial^{4}_{\theta\theta\theta\phi}\right]\\[2.ex]
\end{align*}
\begin{align*}
    P_1=&\, \alpha^2\,(-12\,\hat{H}+8\,\hat{H}^2-\hat{H}^3)+\left[-36+360\,\alpha^2+(120-351\,\alpha^2)\,\hat{H}+(27\,\alpha^2-35)\,\hat{H}^2\right.\\
    +&\left.\,\hat{H}^3\right]\,\hat{H}_\phi+\left[-1737+2511\,\alpha^2+(837-243\,\alpha^2)\hat{H}-27\,\hat{H}^2\right]\,\hat{H}^{2}_{\phi}+\left(-4698+729\,\alpha^2\right.\\
    +&\,\left. 243\,\hat{H}\right)\,\hat{H}^{3}_{\phi}-729\,\hat{H}^{4}_{\phi}\\[2.ex]
    P_2=&\, 36\,\alpha^2+(12-108\,\alpha^2)\,\hat{H}+(27\,\alpha^2-8)\,\hat{H}^2+\hat{H}^3+\left[-396+1377\,\alpha^2\right.\\
    +&\,\left.(459-486\,\alpha^2)\hat{H}-54\,\hat{H}^2\right]\,\hat{H}_\phi+\left[-3888+2187\,\alpha^2+729\,\hat{H}\right]\,\hat{H}^{2}_{\phi}-2916\,\hat{H}^{3}_{\phi}
\end{align*}
\end{itemize}

\section{Conclusions}

This paper has been devoted to the algebraic structure of the
`fundamental symmetries' and their relation to polynomial symmetries
for an example in the class of Lissajous systems. Explicit
expressions of such symmetries as well as their algebraic structure
have been obtained. In this process, we have applied the same method
for both classical and quantum cases, showing the close similarities
of the corresponding expressions and properties. We have chosen a
very simple system to show clearly the main steps of our method, but
it can be applied to a wide class of superintegrable systems. In
this respect, work is in progress to prepare a systematic list of
applications.

Next, we will briefly comment on the results and methods of some
previous references dealing with symmetries of similar systems and
the connection with our results. As it was mentioned in
Lissajous--1, the Lissajous systems include the so called TTW system
\cite{ttw09,ttw10} and a list of similar systems considered in
Refs.\ \cite{post10,post12}. The classical superintegrability of
such systems was proved in many references \cite{pogosyan11, kkm10,
post12,kkm12,kkm12b,gonera12,hakobyan12}. In general, the methods
followed in these references were based on the solutions of the
Hamilton-Jacobi equation  or on action-angle variables arguments.
The superintegrability of the quantum versions of many of these
systems was studied by means of the recurrence properties of the
solutions of the corresponding stationary Schr\"odinger equations
and other techniques. These recurrence relations were applied to get
polynomial symmetry operators \cite{kkm10b,kkm11,kkm13}.

As we have remarked in Lissajous--1 our method implement parallel
procedures for classical and quantum systems, so it is essentially
different from those of the above mentioned references. Our way to
get symmetries is based on algebraic properties of the Hamiltonians
as differential operators (or functions), in other words, on the
existence of shift and ladder  operators (or functions).
Therefore, we have not used the explicit form of any solution to get symmetries
but the other way round.

One of our main objectives has been to know in easy terms the origin
of the algebraic relations for the classical and quantum polynomial symmetries.
Besides this, in our approach
one can directly appreciate how close
are the expressions and properties of classical and quantum systems, see for
instance (\ref{pbps}) and (\ref{pbpsb}).
Finally, we have obtained the Hermitian relations of our polynomial
symmetry algebras,
which are consistent with the commutation relations.

Other references have considered the problem of superintegrability
of classical systems on constant curvature spaces
\cite{ballesteros,ballesteros13} from a different point of view.
Some other methods have been applied to display superintegrability
properties, for instance in classical systems a kind of complex
functions (similar to our shift and ladder functions) have been used
in \cite{ranada12,ranada13}. A coalgebra procedure has been shown to
be useful to find the symmetries of higher dimensional systems
\cite{danilo13}. Other algebraic methods have also been  developed
in \cite{quesne10}.

\subsection*{ {Acknowledgments}}
We acknowledge financial support from GIR of Mathematical Physics of
the University of Valladolid.
\c{S}.~Kuru acknowledges the warm hospitality at Department of
Theoretical Physics, University of Valladolid, Spain.



\end{document}